\documentstyle[12pt]{article}

\begin{document}
\begin{titlepage}
\title{\bf BOUNDARY VALUE PROBLEMS FOR INTEGRABLE EQUATIONS COMPATIBLE
WITH THE SYMMETRY ALGEBRA}
\vspace{5cm}
\author{Burak G{\" u}rel$^{a}$  , Metin G{\" u}rses$^{a}$
and Ismagil Habibullin$^{b}$ \\
{a:\small Department of Mathematics, Faculty of Science} \\
{\small Bilkent University, 06533 Ankara, Turkey}\\
{b:\small Mathematical Institute, Ufa Scientific Center} \\
{\small Russian Academy of Sciences,  Chernishevski str. 112}\\
{\small  Ufa, 450000, Russia}}

\maketitle
\begin{abstract}
Boundary value problems for integrable nonlinear partial differential equations
are considered from the symmetry point of view. Families of boundary
conditions compatible with the Harry-Dym, KdV and MKdV equations
and the Volterra chain are discussed . We also discuss the uniqueness
of some of these boundary conditions.
\end{abstract}
\end{titlepage}

\section{Introduction}
In our previous paper [1] we have briefly discussed a method
to construct boundary value problems of the form

\begin{equation}
u_{t}=f(u,u_{1},u_{2},...,u_{n}) , \label{E01}
\end{equation}

\begin{equation}
p(u,u_{1},u_{2},...,u_{k}) \vert_{x=0}=0 , \label{EE2}
\end{equation}

\noindent
completely compatible with the integrability property of Eq.(1).
Here $u=u(x,t)$ , $u_{i}={\partial^{i}\,u \over \partial\,x^{i}}$ and
$f$ is a scalar (or vector) field.
The aim of the present paper is to expound detailly our scheme and
also extend it to the integrable differential-difference equations.

Let the equation

\begin{equation}
u_{\tau}=g(u,u_{1},...,u_{m}) , \label{EE4}
\end{equation}

\noindent
for a fixed value of $m$, be a symmetry of the equation (\ref{E01}).
Let us introduce some new set of dynamical variables,
consisting of the variable $v=(u,u_{1},u_{2}, ... u_{n-1})$,
and its $t$-derivatives $v_{t}$, $v_{tt}$, ... .
One can express the higher $x$-derivatives of $u$, i.e., $u_{i}$ for
 $i \ge n$
and their $t$-derivatives, by the utility of the equation (\ref{E01}),
in terms of the dynamical variable $v$
and their $t$-derivatives. Here $n$
is the order  of the equation (\ref{E01}). In these terms the symmetry
(\ref{EE4}) may be written as

\begin{equation}
v_{\tau}=G\,(v,v_{t} ,
v_{tt} ,... v_{tt...t}).              \label{EE5}
\end{equation}

\noindent
We call the boundary value problem, Eqs.(\ref{E01}) and (\ref{EE2}),
as compatible with symmetry (\ref{EE4}) if the constraint $p(v)=0$
(or constraints $p^{a}(v)=0$
where $a=1,2,...N$  and $N$ is the number of constraints) is consistent
with the $\tau$-evolution

\begin{equation}
{\partial p \over \partial \tau}=0,\quad (mod\quad p=0)
                                                        \label{EE6}
\end{equation}

\noindent
Eq.(\ref{EE6}), by virtue of the equations in (\ref{EE5}),
must be automatically satisfied. In fact (\ref{EE6}) means that
the constraint
$p=0$ defines an invariant surface in the manifold with
local coordinates $v$. This definition of consistency
of boundary value problem with symmetry is closer to the one introduced
in [3], but not identical. For instance, let us examine whether  the
boundary value problem $u_{t}=u_{xx}$; $u_{x}=c\,u,$ ${x=0},$
is compatible with the symmetry $u_{\tau}=u_{xxx}$. To this end one
has to check if the equation $w=c\,u$ defines an invariant surface
for the system of equations $u_{\tau}=w_{t}$ , $w_{\tau}=u_{tt}$ (here
$w$ is $u_{1}$). Evidently the answer is negative. To check the validity
of compatibility condition in the sense of [3] one has to compare two sets
of equations $u_{2n+1}=c\,u_{2n}$ and $u_{3n+1}=c\,u_{3n}$ , $n \ge 0$.
These equations are obtained by differentiation of the constraint equation
$u_{x}=c\,u$ with respect to $t$ and $\tau$
variables respectively. In this sense the boundary condition is compatible
with the symmetry because the sets of equations don't contradict each other.

We call the boundary condition (\ref{EE2})
is compatible with the equation if it is compatible at least with
one of its higher order symmetries.

Our main observation is that if  the boundary condition is compatible
with one higher symmetry then it is compatible with infinite
number of symmetries. We define a set $S$ with infinite number of elements
where its elements are symmetries of the Eq.(\ref{E01}). $S$ may or may
not contain the whole symmetries of (\ref{E01}). For instance , $S$ contains
the even numbered time independent symmetries for the Burgers equation.

We note that all the known boundary conditions of the form (\ref{EE2})
consistent with the inverse scattering method are indeed compatible with
the infinite series of generalized symmetries. On the other hand,
stationary solutions of the symmetries compatible with (\ref{EE2})
allow one to construct an infinite dimensional set of "exact" (finite gap)
solutions of the corresponding boundary value problem (\ref{E01})
and (\ref{EE2}). However , in this work we do not discuss analytical
aspects of this problem. We note also that , in this paper we shall deal
with boundary conditions of the form given in (\ref{EE2}). An effective
investigation of boundary conditions involving an explicit $t$-dependence
is essentialy more complicated. Such a problem has been studied
 , for instance , in \cite{FO1}.

The article is orginized as follows. In Section 2 we introduce
the necessary notations and give the Burgers equation
as an illustrative example. We prove that if a boundary condition
is compatible at least with one higher order symmetry then it
is compatible with every one of even order. In Section.3 we consider
the nonlinear Schr{\" o}dinger, Harry-Dym, Korteweg de Vries and modified
KdV equations.
Using the symmetry approach we find a boundary conditon compatible
with the symmetry algebra of the Harry-Dym equation

\begin{equation}
\begin{array}{ll}
u_{t}=u^{3}\,u_{xxx} ,\\
u_{x}=c\,u ~~,~~x=0\\
u_{xx}=c^{2}\,u/2 ~~,~~x=0 , \label{EE3}
\end{array}
\end{equation}

\noindent
where $c$ is an arbitrary real constant. Actually one has here two constraints.
Although we are taking the
boundary conditions at $x=0$ , one can shift this point to an arbitrary
point $x=x_{0}$ without loosing any generality. We conjecture
that the boundary value problem given in (\ref{EE3}) is compatible
with the Hamiltonian integrability and solvable by the inverse
scattering technique. In addition we conjecture that (using the idea
in [2]) one can prove that on the finite interval $x_{1} \le x \le x_{2}$
the Harry-Dym equation with the boundary conditions $u_{x}=c_{0}\,u ,
u_{xx}=c_{0}^{2}\,u/2$ for $x=x_{1}$ and $u_{x}=c_{1}\,u ,
u_{xx}=c_{1}^{2}\,u/2$ for $x=x_{2}$ is a completely integrable Hamiltonian
system.

Section 4 is devoted to the differential-difference equations. In the last
section we propose further generalization of the compatibility and discuss
on some open questions.

\section{Boundary Conditions Compatible with Symmetries}

In the sequel we suppose that eq.(\ref{E01}) admits a recursion operator
of the form
(see [4-6])

\begin{equation}
R=\displaystyle \sum_{i=0}^{i_{1}} \alpha_{i}\, D^{i} +
\displaystyle \sum_{i=0}^{k_{1}} \alpha_{-1,i}\,D^{-1}
\alpha_{-2,i} ~~,~~i_{1} \ge 0   ~~,~~k_{1} \ge 0
                                                             \label{EE7}
\end{equation}

\noindent
where $\alpha_{i}$, $\alpha_{-1,i}$, $\alpha_{-2,i}$ are functions of the
dynamical variables, $D$
is the total derivative with respect to $x$. Recursion operators
when applied to a symmetry produce new symmetries. Passing to the new
dynamical variables $v, v_t, v_{tt},...$ one can obtain, from (\ref{EE7}),
the recursion operator of the system of equations (\ref{EE5}) (we don't
prove that every recursion operator may be rewritten in the matrix form, but
we will give below the matrix forms of the recursion operators for the Burgers,
KdV, MKdV and Harry-Dym equations)

\begin{equation}
{\bf R}=\displaystyle \sum_{i=0}^{M}\,a_{i}\,
 (\partial_t)^i
+\displaystyle \sum_{i=0}^{K}\,
a_{-1,i}\, (\partial_t^{-1}) \,a_{-2,i}\, ,\quad M>0, \quad K\ge 0,
                                                                  \label{EE8}
\end{equation}

\noindent
where $a_{i}$ depends on $v$ and on a finite number
of its $t$-derivatives , $\partial_t$ is the operator of the total derivative
with respect to $t$.  If (\ref{E01}) is
a scalar equation, $R$ is a scalar operator, then ${\bf R}$
is an $n \times n$ matrix valued operator.
Our further considerations are based on the following
proposition, which really affirms that if an equation admits
an invariant surface, then a kind of its higher symmetries
admits also the same invariant surface.

{\bf Proposition 2.1.} Let the equation (\ref{EE5}) is of the
form $ v_{\tau}=H({\bf R})v_t$, where ${\bf R}$ is the recursion
operator (8) and $H$ is a polynomial function with scalar constant
coefficients. If this equation is consistent with the constraint
$p(v)=0$, where rank of $p$ equals $n-1$ (here $n$ is the
dimension of the vector $v$) then every equation of the form
$ v_{\tau}=L(H({\bf R}))v_t$, where $L$ is arbitrary chosen scalar
polynomial with constant coefficients, is also compatible
with this constraint.

{\bf Proof:} Introduce new variables $w=(w^1, w^2,...w^n)$
in the following way: $w^1=p^1, w^2=p^2,...w^{n-1}=p^{n-1}$
and $w^n=p^n$ is a function of $v$, here $p^i$ is a coordinate
of the vector $p$. Then
one obtains the equation $w_\tau=Pw_t$ from (5), where $P=A^{-1}
H({\bf R})A$ and $A=\partial v/\partial w$ is the Jacobi matrix of the
mapping $w\rightarrow v$. Notice that under this change of variables
the constraint $p(v)=0$ turns into the equation $w^i=0$ for
$i=1,2,...n-1.$ Imposing this constraint reduces the equation
$w_\tau=Pw_t$ to the form
$$
\pmatrix{0 \cr ...\cr 0 \cr w^n_\tau}=\pmatrix{P_{11}&...& P_{1n}\cr
...&...&...\cr
P_{n-1,1}&...&P_{n-1,n}\cr
P_{n,1}&...&P_{n,n}} \pmatrix{0 \cr ...\cr 0 \cr w^n_t}
$$
Let us show that elements of the last column of the matrix $P$ are
equal zero except maybe $P_{n,n}$: $P_{i,n}=0$ for
$1\leq i \leq n-1$. Really, letting $P_{j,n}\neq 0$
for some $j\leq n-1$ the equation
$P_{j,n} w^n_t=0$ gives a connection between variables
$w^n,\,w^n_t,\,... $ which are supposed to be independent.
Since the set of such block triangular matrices constitutes
a subalgebra in the algebra of the all squared matrices
hence one can easily conclude that the operator
$L(P)(mod w^i=0, i\leq n-1)$ is also block triangular so
the equation $w_\tau=L(P)w_t$ is consistent with the
constraint $w^i=0, i\leq n-1)$. It completes the  proof of
the Proposition 2.1.

Proving later a kind of uniquness theorem (see below the
Proposition 2.2) we will use
the following statement

{\bf Proposition 2.1$^{\prime}$:} Suppose that the constraint
$p(v)=0$ of the rank equal $n-1$ is compatible with the equation
(5) having the form $ v_{\tau}=H({\bf R}^{n_0})v_t$, where $H=H(z)$
is a scalar polynomial function of $z$ with constant coefficients.
Assume that $n_0\geq 1$ is an integer and the leading term
$b_N$ (the coefficient before the highest derivative) in the expression
${\bf R}^{n_0}=b_N(\partial_t)^N+b_{N-1}(\partial_t)^{N-1}+...$
is a scalar matrix, i.e. it is proportional the unit matrix.
Then the constraint is consistent with the equation

\begin{equation}
v_{\tau}={\bf R}^{n_0} \, v_{t}
                                                                   \label{EE9}
\end{equation}

{\bf Proof:} In terms of the variable $w$ we have introduced proving
the previous
proposition the equation $ v_{\tau}=H({\bf R}^{n_0})v_t$ takes the form
$ w_{\tau}=H(R_1^{n_0})w_t$. Owing the fact that the point
transformation preserves the commutativity property of flows ,
the operator $R_1=A{\bf R}A^{-1}$ is the recursion operator in the new
variables. Again, just in the previous proposition one has
that the operator $P=H(R_1^{n_0})$ under the substitution $p=0$
(or really, $w^i=0,i\leq n-1$)
is a lower block triangular matrix valued one. Our aim now is
to prove that the operator $Q =R_1^{n_0} (mod w^i=0,i\leq n-1)$
is also block triangular. Setting $H(Q)=\alpha_nQ^n+
\alpha_{n-1}Q^{n-1}+...+\alpha_0$ and representing $Q$ as formal
series $\sum_{k=-\infty}^{M} c_k\partial_t^k$ using the famous
Campbell-Hausdorf formula one obtains that
$$
H(Q)=\alpha_n(c_M^n(\partial_t)^{nM}+nc_M^{n-1}c_{M-1}(\partial_t)^
{nM-1}+...)+...+\alpha_0
$$
One has that $H(Q)$ belongs the set $M_-$ consisting of the all
lower block triangular matrices. By looking at the coefficients
of different power of the operator $\partial_t$ one can show
that the matrices $c_i$, $i=M-1,M-2,...$ satisfy the equations
$$
c_M^{n-1}c_i+T_i\in M_-,
$$
where $T_i$ are polynomials with scalar coefficients on
variables $c_{i+1}, c_{i+2},... c_M$ and their derivatives.
So, because of assumptions $c_M=b_M\in M_0$ and $det b_M\neq 0$
it is easy to prove by induction that $c_i\in M_0$ for all $i\leq M$.

For illustration let us give the Burgers equation as an example.
The Burgers equation

\begin{equation}
u_{t}=u_{xx}+2\,u\,u_{x}                                       \label{E14}
\end{equation}

\noindent
which possesses the recursion operator of the form

\begin{equation}
R=D+u+u_{x}\, D^{-1}                                            \label{E15}
\end{equation}

\noindent
(see, for instance, [7]). The simplest symmetry of this equation
is $u_{\tau}=u_{x}$. In terms of the new dynamical variables
this symmetry equation takes the form

\begin{equation}
\begin{array}{ll}
u_{\tau}=u_{1} \\
u_{1,\tau}=u_{t}-2\,u\,u_{1}                                     \label{E16}
\end{array}
\end{equation}

\noindent
This equation does not admit any invariant surface of the
form $p(u,u_{1})=0$. Really, differentiating this constraint with respect to
$\tau$
one obtains

\begin{equation}
{ \partial p \over \partial u}\,u_{1}
+{ \partial p \over \partial u_{1}} \, (u_{t}-2\,u\,u_{1})=0      \label{E17}
\end{equation}

\noindent
Because of independence of the variables $u_{t}$  and $u_{1}$ we have

\begin{equation}
{ \partial p \over \partial u_{1}} = { \partial p \over \partial u}=0
                                                                    \label{E18}
\end{equation}

\noindent
which leads to a trivial solution $p=constant$. As a conclusion
we don't have any invariant surface (curve) in  ($u,u_{1}$) - plane.
Similarly the third order symmetry $u_{\tau}=u_{3}+3uu_{2}+3u_{1}^{2}
+3u^{2}u_{1}$ rewritten in the new variables $(u,u_{1})$ gives
the following system of two equations

\begin{equation}
\begin{array}{ll}
u_{\tau}=u_{1,t}+u\,u_{t}+(u^{2}+u_{1})\,u_{1},\\
u_{1,\tau}=u_{tt}-u\,u_{1,t}+(u^{2}+u_{1})\,u_{t}-
2\,u\,u_{1}\,(u^{2}+u_{1})
\label{E19}
\end{array}
\end{equation}

\noindent
This system also does not admit any invariant surface of the form\\
$p(u,u_{1})=0$. It may be easily proved that the same is true for
every symmetry of the odd order , i.e .,
$u_{\tau}=u_{2m+1}+h(u_{2m},...,u)$. Because the correspondent system
of equations has different orders in the highest $t$-derivatives

\begin{equation}
u_{\tau}=\partial^{m}_{t}\,u_{1}+... ~~,~~
u_{1,\tau}=\partial^{m+1}_{t}\,u+...
\label{E20}
\end{equation}

\noindent
Unlike the symmetries of odd order, for the symmetries of even order
the correspondent system of equations has the same orders
in the highest $t$-derivatives. This fact leads us to show that
the symmetries of even order admit an invariant surface $p(u,u_{1})=0$,
depending upon two arbitrary parameters.

 {\bf Proposition 2.2:} If the boundary condition $p(u,u_{1}) \vert_{x=0}=0$
 is compatible with a higher symmetry of the Burgers equation, then it is
 of the form (see [3]) $c(u_{1}+u^{2})+c_{1}\,u+c_{2}=0$ and is compatible
 with every symmetry of the form $u_{\tau}=P(R^{2})\,u_{t}$  where $P$
 denotes polynomials with scalar constant coefficients.

 {\bf Proof:}The Frechet derivative of (\ref{E14}) gives
 the symmetry equation of the Burgers equation

\begin{equation}
\partial_{t}\, \sigma = (D^{2}+2\,u\,D + 2\,w)\, \sigma
\label{E21}
\end{equation}

\noindent
where $w$ stands for $u_{1}$. As the operators acting on symmetries
we may take

\begin{equation}
D^{-1}=\partial^{-1}_{t}\,(D+2\,u)                                  \label{E22}
\end{equation}

\noindent
in the recursion operator (\ref{E15}). Consequently the recursion
formula $u_{\tau_{i+1}}=R\,u_{\tau_{i}}$ becomes

\begin{equation}
u_{\tau_{i+1}}=(u+2\,w\, \partial^{-1}_{t}\,u)\,u_{\tau_{i}}
+(1+w\, \partial^{-1}_{t}\,)\,w_{\tau_{i}}                          \label{E23}
\end{equation}

\noindent
Differentiating it with respect to $x$ and replacing
$w_{x}=u_{2}=u_{t}-2\,u\,w$ one obtains

\begin{equation}
w_{\tau_{i+1}}=[\partial_{t}+2\,(u_{t}-2\,u\,w)\, \partial^{-1}_{t}\,u]\,
u_{\tau_{i}}+[-u+(u_{t}-2\,u\,w)\, \partial^{-1}_{t}]\,w_{\tau_{i}}
\label{E24}
\end{equation}

\noindent
for $i=1,2,...$. Thus the matrix form of the recursion operator
${\bf R}$ is given by

\begin{equation}
{\bf R}=\pmatrix{u+2\,w\, \partial^{-1}_{t}\,u& 1+w\, \partial^{-1}_{t}\cr
        \partial_{t}+2\,(u_{t}-2\,u\,w)\, \partial^{-1}_{t}\,u&
        -u+(u_{t}-2\,u\,w)\, \partial^{-1}_{t} }
\label{E25}
\end{equation}

\noindent
It is well known that every higher order local polynomial symmetry
may be represented as a polynomial operator $P_{0}(R)$ applied to the
simplest classical symmtery $u_{\tau}=u_{x}$. It is more convenient
to use the following equivalent representation

\begin{equation}
\pmatrix{u \cr
     w}_{\tau}=P({\bf R^{2}})\, \pmatrix{u \cr
                                      w}_{t}
                 +P_{1}({\bf R^{2}})\, \pmatrix{w \cr
                                           u_{t}-2\,u\,w}
                                                                    \label{E26}
\end{equation}

\noindent
where $P$ and $P_{1}$ are polynomials with scalar constnat coefficients and
$P_{0}$ mentioned above may taken as $P_{0}({\bf R})=P({\bf R^{2}})\, {\bf R}
+P_{1}({\bf R^{2}})$.

Note that one could not apply immediately the
Proposition 2.1$^{\prime}$ to this because
the coefficient of $\partial_{t}$  in the representation  (\ref{E25})
is not diagonal. On the other hand the operator ${\bf R^{2}}$
has scalar leading part. First we will prove that if the symmetry
(\ref{E26}) admits an invariant surface then $P_{1}$ in this equation
vanishes. Let us take the invariant surface as $u=q(w)$. Suppose that
the function $q(w)$ is differentiable at some point $w=w_{0}$. Linearizing
$q$ around the point $w_{0}$ (or as $w \rightarrow w_{0}$) we obtain

$$
u-q(w_{0})=q^{\prime}(w_{0})\,(w-w_{0})+o(w-w_{0}).
$$

\noindent
It follows from (\ref{E25}) that in this case ${\bf R^{2}}$
reduces to a scalar operator: ${\bf R^{2}} \rightarrow
(\partial_{t}-w_{0}+q^{2}(w_{0}))\,I$ as $w \rightarrow w_{0}$, where $I$
is the unit matrix. Thus in the linear approximation the Eq.(\ref{E26})
takes the form

\begin{equation}
\pmatrix{u \cr
          w}_{\tau}=P(\partial_{t}-w_{0}+q^{2}(w_{0}))\,
       \, \pmatrix{u \cr
                  w}_{t}+ P_{1}(\partial_{t}-w_{0}+q^{2}(w_{0}))\,
                                 \, \pmatrix{w \cr
                                            u_{t}}
\label{E27}
\end{equation}

\noindent
where now $P(\partial_{t}-w_{0}+q^{2}(w_{0}))\,$ and
$P_1 (\partial_{t}-w_{0}+q^{2}(w_{0}))$ are scalar operators. It is
clear that the linearized equation is consistent with the
linearized boundary condition $ u-q(w_{0})=q^{\prime}(w_{0})\,(w-w_{0}) $ ,
provided $P_{1}=0$. Supposing that the equation (\ref{E26}) is compatible
with the constraint $w=c$ where $c$ is a constant and then linearizing
about the point ($u=0,w=c$) one can easily obtain that $P_{1}$ vanishes
in this case also.

It is evident now that in Proposition 2.1$^{\prime}$
one should put $n_0=2,$ because
 ${\bf R^{2}}=I\, \partial_{t} + ...$. With this choice the
constraint $p(u,w)$ describes an invariant surface for the following system

\begin{equation}
\pmatrix{u \cr w}_{\tau}={\bf R^{2}}\, \pmatrix{u \cr w}_{t}
\label{E28}
\end{equation}

\noindent
which is exactly the coupled Burgers type integrable system
(see [8])

\begin{equation}
\begin{array}{ll}
u_{\tau}=u_{tt}+2(w+u^{2})\,u_{t} \\
w_{\tau}=w_{tt}+2u_{t}^{2}+2(w+u^{2})\,w_{t}
\label{E29}
\end{array}
\end{equation}

\noindent
It is straightforward to show that the above system (\ref{E29})
is compatible with the constraint $p(u,w)=0$ only if $p=w+u^{2}+c_{1}\,u
+c_{2}$ or $u=const$.

The above uniqeness proof of the boundary condition
$p=w+u^{2}+c_{1}\,u+c_{2}$ can be more easily shown
if we use a new property of the Burger's hierarchy.
We have the following propositon:

\noindent
{\bf Propositon 2.3:}
The function $u(t,x,\tau_{n})$  ($ n\ge 1$) satisfy infinitely many
Burgers like equations

\begin{equation}
u_{,\tau_{i},\tau_{i}} - u_{,\tau_{2i+2}} =
-2\, u_{,\tau_{i}}\,D^{-1}\,u_{,\tau_{i}} \label{s11}
\end{equation}

\noindent
Here $i=-1,0,1,2,...$. Burgers equation corresponds to $i=-1$ ,
$\tau_{-1}=x$ , and $\tau_{0}=t$. All  $u_{\tau_{i}}$ for $i>-1$ correspond to
higher
symmetries. It is straightforward to determine the even numbered
symmetries of the Burgers equation from (\ref{s11}). It is very
interesting that $u$ satisfies the Burgers like equations
with respect to the variables ($\tau_{i}$ , $\tau_{2i+2}$)
for all $i=-1,0,1,2,...$ .

The proof of this proposition depend crucially on definition
of the higher symmetries of the Burgers equation.
They are defined through the equation

\begin{equation}
u_{\tau_{n}}=R^{n+1}\,u_{x}
\label{t11}
\end{equation}

\noindent
where $R$ is the recursion operator given in eq.(\ref{E15}) and $n \ge -1$.
Eq.(\ref{t11}) can also be written as $u_{\tau_{n}}=R\,u_{\tau_{n-1}}$.
Differentiating this equation once by $\tau_{n}$
and using (\ref{t11}) one arrives at (\ref{s11}).

If we let the most general boundary condition of the form
$p=f(u,u_{x})= 0$ at $x=x_{0}$ and take $\tau_{i}$ and $\tau_{2i+2}$
 derivatives (for $i \ge 0$) of the function $p$ and use the equation
 (\ref{s11}) we obtain

\begin{equation}
f_{u_{x}}^{2}\,f_{,u,u}+f_{u}^{2}\,f_{,u_{x},u_{x}}-2\,f_{,u_{x}}^{3}
-2\,f_{,u}\,f_{,u_{x}}\,f_{u,u_{x}}=0
\label{s12}
\end{equation}

\noindent
Letting $u=x_{1}$ and $u_{x}+u^{2}+c_{1}\,u+c_{2}=x_{2}$ then
eq.(\ref{s12}) becomes

\begin{equation}
f_{,x_{2}}^{2}\,f_{,x_{1},x_{1}}+f_{,x_{1}}^{2}\,f_{,x_{2},x_{2}}
-2\,f_{,x_{1}}\,f_{,x_{2}}\,f_{,x_{1},x_{2}}=0 ,\label{s13}
\end{equation}

Assuming $f_{x_{2}} \ne 0$ and letting $q=f_{,x_{1}}/f_{,x_{2}}$
we find that

\begin{equation}
q_{,x_{1}}=q\,q_{,x_{2}}
\label{s14}
\end{equation}

\noindent
This is a very simple equation and its general solution can be found.
We shall not follow this direction to determine $f(x_{1}, x_{2})$ rather
change the form of equation $p(u,u_{x})=0$ at $x=x_{0}$. This equation
(in principle) implies either

\noindent
{\bf a)}\, $u_{x}=h(u)$

\noindent
which implies $f=u_{x}-h(u)$ at $x=x_{0}$ . Or

\noindent
{\bf b)}\, $u=g(u_{x})$

\noindent
which implies $f=u-g(u_{x})$ at $x=x_{0}$
It is now very easy to show that the cases {\bf a} and {\bf b}
when the corresponding $f$'s are inserted in (\ref{s12}) we
respectively obtain

\noindent
{\bf a)}\, $h^{''}+2=0$

\noindent
which implies  $u_{x}+u^{2}+c_{1}\,u+c_{2}=0$ at $x=x_{0}$

\noindent
{\bf b)}\, $g^{''}+2\,(g^{'})^{3}=0$

\noindent
which implies $u=constant$ (for $g^{'}=0$) and a special case of {\bf a}
(for $g^{'} \ne 0$). Hence we found all possible bounday conditions.

{\bf Remark 2.3:} On the invariant surface $p(u,w)=0$
the system (\ref{E29}) turns into the Burgers like equation
$u_{\tau}=u_{tt}-2(c_1\,u+c_{2})\,u_t$ which is also integrable [5].

\section{Application to Other Partial Differential Equations}

In this section we shall apply our method to obtain compatible
boundary conditions of some nonlinear partial differential equations.
Let us start with the following system of equations

\begin{equation}
\begin{array}{ll}
u_{t}=u_{2}+2u^{2}\,v \\
-v_{t}=v_{2}+2u\,v^{2}
\end{array}
\end{equation}

\noindent
Letting $v=u^*,$ $t \rightarrow i\,t$ the above system becomes
the well known
nonlinear Schr{\" o}dinger equation. Suppose that it admits a
boundary condition of the following form

\begin{equation}
u_{x} \vert_{x=0}=p^{1}(u,v) ~~,~~ v_{x} \vert_{x=0}=p^{2}(u,v)
\label{E30}
\end{equation}

\noindent
compatible with the fourth order symmetry. It means that
the constraint (\ref{E30}) defines an invariant surface for this symmetry,
presented as a system of four equations with two
independent variables

\begin{equation}
\begin{array}{ll}
u_{\tau}=u_{tt}-2u^{2}v_{t}-4u\,v_{1}\,u_{1}+2\,v\,u_{1}^{2}-2\,u^{3}\,v^{2}
,\\
v_{\tau}=-v_{tt}-2\,v^{2}\,u_{t}+4\,v\,u_{1}\,v_{1}-2\,u\,v_{1}^{2}+2\,v^{3}\,u^{2} ,\\
u_{1,\tau}=u_{1,tt}-2u^2v_{1t}    -2\,u_{1}^{2}\,v_{1}-6\,u^{2}\,v^{2}\,u_{1}-
4\,u\,v_{1}\,u_{t}+4\,v\,u_{1}\,u_{t}+4\,v\,u^{3}\,v_{1},\\
v_{1,\tau}=-v_{1,tt}-2\,v^{2}\,u_{1,t}+2\,v_{1}^{2}\,u_{1}+6\,v_{1}\,v^{2}\,
u^{2}-4\,v\,u_{1}\,v_{t}+4\,u\,v_{1}\,v_{t}-4\,v^{3}\,u\,u_{1}
\label{E31}
\end{array}
\end{equation}

\noindent
One can check that the system (\ref{E31}) is compatible with the constraint
$u_{1}=p^{1}(u,v) , v_{1}=p^{2}(u,v)$ only if $p^{1}=c\,u$ and
$p^{2}=c\,v$. Since the system (\ref{E31}) is of the form

\begin{equation}
(u,u_1,v,v_1)^{T}_{\tau}={\bf R^{2}}\,(u,u_1,v,v_1)_{t}^T
\end{equation}

\noindent
hence it follows from the Proposition 2.1 that the constraints
$u_{1}=c\,u$, $v_{1}=c\,v$ are compatible with every symmetry of even order.
So the boundary conditions $u_{x} \vert_{x=0}=c\,u$ ,
$v_{x} \vert_{x=0}=c\,v$ are compatible with such symmetries. Analytical
properties of this boundary value problem are studied previously
(see [2],[9-10]) by means of the inverse scattering method.

{\bf Remark 3.1:} On the invariant surface $u_{1}=c\,u~ ,~ v_{1}=c\,v$
the system (\ref{E31}) is reduced to a system of two equations:
$$
u_{\tau}=u_{tt}-2u^{2}v_{t}-2c^2 u^2 \,v -2\,u^{3}\,v^{2} ,
$$
$$
v_{\tau}=-v_{tt}-2\,v^{2}\,u_{t}+2c^2\,v^2\,u+2\,v^{3}\,u^{2} ,
$$
\noindent
The integrability of these equations is shown in [4] (see p.175).
Under a suitable change of variables in it this system of two equations
becomes the famous derivative nonlinear Schr{\" o}dinger equation.

Among the nonlinear integrable equations the Harry-Dym equation

\begin{equation}
u_{t}+u^{3}\,u_{3}=0
\label{E32}
\end{equation}

\noindent
is of special interest. It is not  quasilinear and because of this reason
its analytical properties are not typical. Using the symmetry approach
we find a boundary condition of the form

\begin{equation}
p(u,u_{1},u_{2})=0 ,                                              \label{E33}
\end{equation}

\noindent
compatible with Harry-Dym equation. One has to notice that because of
non-quasilinearity of (\ref{E32}) the transformation from the standard
set of variables $u,u_{1},u_{2},u_3...$ to
$u,u_{1},u_{2},u_{t},u_{1,t},u_{2,t},...$ is not regular. For instance
$u_{3}=-{u_{t} \over u^3}$. It has singular
surface given by the equation $u=0$. So one should examine
this surface separately. Since the Harry-Dym equation (\ref{E32})
as well as its higher order symmetries possesses the reflection
symmetry $x \rightarrow -x , u \rightarrow -u , t \rightarrow t $
the trivial boundary condition $u(t,0)=0$ is consistent with
the integrability.

Suppose that the boundary value problem (\ref{E32}) and (\ref{E33})
is compatible with the ninth order symmtery $u_{\tau}=u^{9}\,u_{9}+...$.
It means that the constraint $p(u,v,w)$ is consistent with following
system of equations , equivalent to the ninth symmetry

\begin{equation}
\begin{array}{ll}
u_{\tau}=f_{1}\\
v_{\tau}=f_{2}\\
w_{\tau}=f_{3}   \label{ss1}
\end{array}
\end{equation}

\noindent
where $v=u_{x},$ $w=u_{xx}$ and $(f_1,f_2,f_3)^T={\bf R^3}(u_t,v_t,w_t)^T,$
$$
{\bf R}=\pmatrix{uw+u_t\, \partial^{-1}_{t}\,w&
                        -uv-u_t\, \partial^{-1}_{t}v&
                             u^2+u_t \, \partial^{-1}_{t}u
 \cr
{1\over u}\,\partial_t+vw-{{u_t}\over {u^2}}+v_t\,\partial_t^{-1}w &
                                     -v^2-v_t\,\partial_{t}^{-1}v&
                                         uv+v_t\, \partial^{-1}_{t}u
  \cr
w^2+w_t\,\partial_t^{-1}w&
  {1\over u}\partial_t\,-vw-{{u_t}\over {u^2}}-w_t\,\partial_t^{-1}v&
                                            uw+w_t\,\partial_t^{-1}u}.$$

\noindent
The explicit expressions for $f_2$, $f_3$ are very long. Hence we give
the explicit form only for the function $f_1$:
\begin{equation}
\begin{array}{ll}
f_1=-u_{ttt}+3u_{tt}u_t{1\over u}-{3\over2} u_{tt}u_1h-
{3\over2} {u_t^3\over u^2} +{3\over 2}u u_{1,tt}h+\\+{3\over 2}uu_{1,t}h_t-
{{15}\over
{16}}uh^2h_t-{5\over{16}}h^3u_t-{3\over2}u_1u_th_t .
\label{EEE}
\end{array}
\end{equation}

Where $h=2u_2u-u_1^2$. Here one has two choices for the rank of the
equation (\ref{E33}). It is either one or two. The first choice does not
lead to any regular invariant surface. The second gives

\begin{equation}
u_{x} \vert_{x=0}=c\,u ~~,~~ u_{xx} \vert_{x=0}={c^{2}\, u \over 2}
\label{E34}
\end{equation}
{\bf Remark 3.3 :} On the invariant surface $v=cu,$ $w=c^2u/2$ the first
equation in the system
(\ref{ss1}) takes the form

\begin{equation}
u_{\tau}=-u_{ttt}+3u_tu_{tt}/u -3u_t^3u^2/2
\label{ss2}
\end{equation}

\noindent
equivalent to the MKdV equation.

Since the symmetry under consideration is of the form
$u_{\tau}=R^{3}\,u_{x}$ where $R=u^{3}\,D^{3}\,u\,D^{-1}\, {1 \over u^{2}}$
the recursion operator for the Harry-Dym equation (see [11]) , the Propositon
2.1  implies the following

{\bf Proposition 3.1:} The boundary value problem  (\ref{E32}) and (\ref{E33})
is compatible with every symmetry of the form $u_{\tau}= L(R^{3})\,u_{x}$ ,
where $L$ is a polynomial with scalar constant coefficients.

The Korteweg de Vries equation $u_t=u_{xxx}+6u_1u$ admits a recursion
operator $R=D^2+4u+2u_1D^{-1}$ which may be represented in the form:
$$
{\bf R}=\pmatrix{4u +12v\, \partial^{-1}_{t}\,u&
                             0&
                                     1+2v \, \partial^{-1}_{t}
 \cr
\partial_t+12w\,\partial_t^{-1}u&
                                     -2u&
                                         2w\, \partial^{-1}_{t}
  \cr
2w+12(u_t- 6uv)\,\partial_t^{-1}u&
                      \partial_t\,-2v&
                                        -2u +2(u_t-6uv)\,\partial_t^{-1}}
.$$
It is not difficult to show that the system of equations
$(u,v,w)_{\tau}={\bf R}^3(u,v,w)_t$ admits an invariant surface
$u=0,$ $w=0$ on which the equation turns into the MKdV equation. It means that
the boundary condition $u(t,x=0)=0,$ $u_{xx}(t,x=0)=0$ is compatible with
all symmetries of the form $u_{\tau}=R^{3n}u_x.$
Similarly, the MKdV equation $u_t=u_{xxx}+6u^2u_{x}$ is compatible with the
boundary condition $u(t,x=0)=0,$ $u_x(t,x=0)=0.$

\section{Application to Discrete Chains}

Consider an integrable nonlinear chain of the form
\begin{equation}
u_{t}(n)=f(u(n-1), u(n), u(n+1))
\label{4E1}
\end{equation}

\noindent
with unknown function $u=u(n,t)$ depending on integer $n$ and
real $t$. The natural set of dynamical variables serving the hierarchy
of higher symmetries for the chain is the set
$u(0), u(\pm 1), u(\pm 2), ...$\,.
However, it is more convenient for our aim to use the following
unusual one, consisting of the variables $u(0), u(1)$ and all their
$t$-derivatives. Transformations of these sets to each other are given
by the equation (\ref {4E2}) itself and its differential consequences. In terms
of new basic variables every higher order symmetry of this chain

\begin{equation}
u_{\tau}(n)=g(u(n-m),u(n-m-1),...u(n+m)) , \label{4E2}
\end{equation}

\noindent
could be presented as a system of two partial differential equations

\begin{equation}
\begin{array}{ll}
v_{\tau}=G_{1}(v,w,v_{1},w_{1}...,v_{s},w_{s}) , \\
w_{\tau}=G_{2}(v,w,v_{1},w_{1}...,v_{s},w_{s}) , \label{4E3}
\end{array}
\end{equation}

\noindent
where $v=u(0,t,\tau )$, $w=u(1,t,\tau)$, $v_{i}={\partial^{i}\,v \over
\partial\,t^{i}}$,  $w_{i}={\partial^{i}\,w \over \partial\,x^{i}}.$
\par
Prescribe some boundary condition of the form
\begin{equation}
u(0)=p(u(1),u(2),...u(k))    \label{4E4}
\end{equation}

\noindent
to the equation (\ref {4E1}) to hold for all moments $t.$
We shall call the boundary value problem  (\ref {4E1}), (\ref {4E4})
consistent with the
symmetry (\ref {4E2}) if the constraint (\ref {4E4}) defines an invariant
 surface
for the system (\ref {4E3}).
        Note that interconnection between the hierarchies of the commuting
discrete chains and integrable partial differential equations is well-known
(see survey [4]). An illustrative example of such a kind connection
is related to the famous Volterra chain
\begin{equation}
u_{t}(n)=u(n)(u(n+1)-u(n-1))
\label{4E5}
\end{equation}
\noindent
for which the next symmetry

$$ u_{\tau}(n)=u(n)u(n+1)(u(n)+u(n+1)\\+u(n+2))-$$
$$u(n)u(n-1)(u(n)+u(n-1)+u(n-2))$$

\noindent
which might be represented as ([4], p.123)

\begin{equation}
\begin{array}{ll}
v_{\tau}+ v_{tt}=(2vw+v^2)_{t},\\
w_{\tau}- w_{tt}=(2vw+w^2)_{t}
\label{4E6}
\end{array}
\end{equation}

\noindent
under the substitution $u(0)=v,$  $u(1)=w,$ $u(-1)=w-{{v_{t}}\over {v}},$\\
$u(2)=v+{{w_{t}}\over {w}},$
$u(-2)=v-{\partial \ln u(-1) \over \partial t}.$ Moreover, the full hierarchy
of the Volterra chain is completely described by the hierarchy of the last
system.
According to the definition above the boundary value problem (\ref{4E4}),
(\ref{4E5}) will
be consistent with a symmetry of the Volterra chain if the constraint
(\ref{4E4})
describes an invariant surface for the same symmetry, represented as a system
of partial differential equations.
Let us examine invariant surfaces of the following system of partial
differential equations
\begin{equation}
\begin{array}{ll}
v_{\tau}= v_{ttt}+(3vH^2 -3v_{t}H -2v^3)_{t},\\
w_{\tau}= w_{ttt}+(3wH^2 +3w_{t}H -2w^3)_{t},
                                                                \label{4E7}
\end{array}
\end{equation}

\noindent
$H=v+w,$ which is exactly the higher order symmetry for the
Volterra chain (\ref{4E5})
of the form

$$u_{\tau}(n)=u(n)u(n+1)(u(n+2)u(n+3)+u(n)u(n+2)\\+u(n)u(n-1)+$$
$$u^2(n)+2u(n+1)u(n+2)+u^2(n+2)+2u(n)u(n+1)+$$
$$u^2(n+1))-u(n)u(n-1)( u(n)u(n+1)+u(n)u(n-2)+u(n-2)u(n-3)\\+$$
$$u^2(n-2)+2u(n)u(n-1)+u^2(n)+2u(n-1)u(n-2)+u^2(n-1)).$$

\noindent
It is easy to check that the only invariant surface of the form
$v=const$ admissible by the system (\ref{4E4}) is $v=0.$ The corresponding
boundary condition $u(0)=0$ is well studied (see [12], [13]).

{\bf Remark 4.1:} On the invariant surface $v=0$ the system (\ref{4E7})
reduces to the scalar equation
$$
w_{\tau}=w_{ttt}+3w_{tt}w+3w^2_{t} +3w_tw^2,
$$
\noindent
which is nothing else but the next symmetry of the Burgers equation.
Moreover, the constraint is compatible with every generalized polynomial
symmetry. On the invariant surface they are all reduced to the symmetries of
 the Burgers equation. It is evident for instance, that the system
(\ref{4E7}) turns into the Burgers equation itself.
\par
        Suppose now that $v=p(w).$ Then one obtains that $p(w)=-w.$ It
gives rise to a boundary $u(0)=-u(1)$ compatible with the Volterra chain
(see [14]).
\par
{\bf Remark 4.2:} Under the constraint $v=-w$ the system
(\ref{4E7}) turns into
the Modified  KdV equation
$$
v_{\tau}=v_{ttt}+6v^2v_t.
$$
\par
It is not difficult to show that there is no any invariant surface of
the form $v=p(w,w_t)$ such that ${{\partial\, p}\over {\partial\,w_t}}=0$
admissible with the system (\ref{4E7}).
\par
For the case $v_t=p(v,w,w_t)$ calculations become very long so that here
we utilized Matematica 2.1
(we thank G.Alekseev for his help with this calculations). Here $p$ has
a form $p={v\over w } w_t +2v(v+w)$ which produces the boundary condition\\
$u(-1)=-u(0)-u(1)-u(2).$ The slight difference with the (\ref{4E4}) is
overcomed
by the simple shift of the discrete variable $n.$
\par
Using the proposition 2.1 it is easy to check that
the invariant surface\\
$v_t={v\over w } w_t +2v(v+w)$ is compatible with every odd order
polynomial generalized symmetry of the system (\ref{4E7}). It means that
the boundary condition\\ $u(-1)=-u(0)-u(1)-u(2)$ is compatible with the
corresponding symmetries of the Volterra chain.
\par
The well-known boundary condition $u^2(0)=1$ for the modified Volterra
chain

$$u_t(n)=(1-u^2(n))(u(n+1)-u(n-1))$$

\noindent
defines the invariant surface
$v^2=1$ for the following systems of equations
\begin{equation}
\begin{array}{ll}
v_{\tau}+v_{tt}=2((1-v^2)w)_t,\\
w_{\tau}-w_{tt}=2((1-w^2)v)_t
\label{4E8}
\end{array}
\end{equation}
\noindent
and
\begin{equation}
\begin{array}{ll}
v_{\tau}+v_{ttt}=2(v(1-v^2)(3w^2-1)-3vwv_t)_t,\\
w_{\tau}+w_{ttt}=2(w(1-w^2)(3v^2-1)+3vww_t)_t
\label{4E9}
\end{array}
\end{equation}
\noindent
which are equivalent to the next symmetries of this chain:

$$u_{\tau}(n)=(1-u^2(n))(D_- -D_+)(1-u^2(n))(D_- -D_+)u(n)$$
\noindent
and

\noindent
$$u_{\tau}(n)=(1-u^2(n))(D_- -D_+)(1-u^2(n))[(-D^2_+-D^2_-)u(n)+$$
$$(D_++D_-)(u^2(n)u(n+1) +u^2(n)u(n-1)+2u(n-1)u(n)u(n+1)],$$

\noindent
here $D_+,$ $D_-$ are the shift operators: $D_+u(n)=u(n+1),$
$D_-u(n)=u(n-1);$ $v=u(0),$ $w=u(1)$ and other variables $u(n)$ are
expressed through $v,w$ and their $t$-derivatives by means the chain
and its differential consequences.

{\bf Remark 4.3:} On the invariant surface $v^2=1$ the systems (\ref{4E8}),
(\ref{4E9}) are reduced to the Burgers equation and its third order symmetry.

\section{Condition of weak compatibility}

It is easy to notice that any symmetry of the equation
(1.1) rewritten in terms of the non-standard set of the dynamical
variables turns into the equation containing $m-1$ extra variables
$u_1,u_2,...,u_{m-1}$ . For instance, the fourth order symmetry of the Burgers
equation

$$u_{\tau}=u_{4}+4u_3u + 10u_2u_1+6u_2u^2+12u_1^2u +4u_1u^3$$

\noindent
takes the following form

\noindent
$$ u_{\tau}=u_{tt}+2(w+u^2)u_t,$$

\noindent
where $w=u_1.$ To extend it to the closed
form it is enough to add one more equation obtained from the above equation
by the differentiation with respect to $x$ and replacing $u_2=u_t-2uw$.
This is the general rule for integrable equations: one has
to add $m-1$ more equations (to have a closed system of
equations), expressing variables $u_{i\tau},$
$1\leq i \leq m-1$ through dynamical ones.
But from the other hand side one may
consider the single symmetry equation alone and
suppose the extra variables are expressed interms of  $u$ and its lower
derivatives.
Let us pose the question ,
for which choice of such expressions the symmetry
under consideration turns into an integrable equation?
As an example let us consider the Burgers equation
How should we choose the dependence $w=w(u)$, such that the equation
$ u_{\tau}=u_{tt}+2(w+u^2)u_t$ would be integrable? The only choice is
$w=-u^2+c_1u+c_2$ (see [6]). We will call the boundary conditions
$u_i=u_i(u),$ $x=0$ (obtained this way) for the equation (1.1)
as weakly compatible with  the
symmetry if these constraints are chosen to satisfy the requirement
above: i.e. the equation for the $nth$ symmetry written down
in terms of the introduced variables
turns into some integrable equation after replacing $u_i=u_i(u),$
$u_{it}=u_t{{\partial\,u_i}\over {\partial\,u}},... \,.$   So in the above
case of the Burgers equation only the condition $w(u)=-u^2+c_1u+c_2$ is
weakly compatible with the fourth order symmetry. As the remarks
given above indicate,
the compatibility of the condition with a symmetry implies the weak
compatibility with it , but not vice versa. However, we conjecture that
if the boundary condition is weakly compatible with at least
three higher
symmetries then the corresponding initial boundary value problem will be
solvable by a suitable generalization of the inverse scattering method.
\par
The following example for the Harry-Dym equation (\ref{E32}) seems to be
intriguing.
Let us represent the fifth order symmetry

$$u_{\tau_5}=-{ 1\over2} u^3(2u_5u^2+
10u_4u_1u+10u_3u_2u+5u_3u^2_1)$$

\noindent
in the
form $u_{\tau_5}={1 \over 2}\,(hu)_t,$ where $ h=2u_2u-u^2_1.$ Represent also
the next two
symmetries in the similar form:

$$u_{\tau_7}=u_{tt}u_1-{3\over2}u_tu_1uh+
{3\over8}u_t[3(h+u^2_1)^2 -4u^2_1(h+u^2_1)+u^4_1]-uu_{1tt}+{3\over8}u_{2t}uh$$

\noindent
and $u_{\tau_9}=f_1$ (see above the first equation of the system (\ref{ss1})).
It is evident that for arbitrary function $F=F(u)$ the constraint $h=0,$
$u_1=F(u)$ is weakly consistent with fifth and ninth symmetries,
because the former takes the trivial form $u_{\tau_5}=0$ and the latter
turns into the integrable equation (\ref{ss2}). The seventh order symmetry
becomes
$u_{\tau_7}=(Su_t)_t,$ where $S=F-uF'.$ Thus, if for instance, $S=a=const$
or $S={1\over {(\gamma u+\beta)^2}}$ one will have the equation
$u_{\tau_7}=(Su_t)_t,$  to be integrable (see [5], p.129).
 Supposing
$S(u)=a$ one can easily find that $u_1=cu+a,$ $u_2={{c^2u}\over2} +ac +
{{a^2}\over{2u}}.$ It leads to the following boundary condition
$u_x=cu+a,$ $u_{xx}={{u_x^2}\over{2u}},$ $x=0$ for the  Harry-Dym equation,
which coincides with (\ref{E34}) if $a=0.$ In the case
 $S={1\over {(\gamma u+\beta)^2}}$ to find $F$ one has to integrate
the ordinary differential equation
$F(u)-uF'(u)=S.$
\medskip
\par
This work has been supported by the Turkish Scientific and Technical
Research Council (TUBITAK). One of us (I.H.) thanks to the Russian
Foundation for Fundamental Research, grant 93-011-165 for partial
support and Bilkent University for warm hospitality.

\end{document}